\documentclass[12pt]{iopart}
\usepackage{amssymb}
\usepackage{epsfig}

\def\sgn{{\rm sgn}}
\def\be{\begin{equation}}
\def\ee{\end{equation}}
\def\bea{\begin{eqnarray}}
\def\eea{\end{eqnarray}}
\def\bJ{{\bf J}}

\def\bs{{\bf s}}
\def\bsigma{\mbox{\boldmath $\sigma$}}
\def\bpsi{\mbox{\boldmath $\psi$}}
\def\D{\mbox{\rm D}}
\def\d{\mbox{\rm d}}
\def\erf{\mbox{\rm erf}}
\def\lb{\left}
\def\rb{\right}
\def\nn{\nonumber}
\newcommand{\fr}[2]{\frac{#1}{#2}}

\begin{document}
\title{On-line learning and generalisation in coupled perceptrons}
\author{D Boll\'e and P Koz{\l}owski}
\address{Instituut voor Theoretische Fysica, K.U. Leuven, B-3001 Leuven,
Belgium}
\eads{\mailto{desire.bolle@fys.kuleuven.ac.be},
\mailto{piotr.kozlowski@fys.kuleuven.ac.be}}

\begin{abstract}
We study supervised learning and generalisation in coupled perceptrons
trained on-line using two learning scenarios. In the first scenario
the teacher and the student are
independent networks and both are represented by an Ashkin-Teller
perceptron. In the second scenario the student and the teacher are simple
perceptrons but are coupled by an Ashkin-Teller type four-neuron interaction
term.
Expressions for the generalisation error and the learning curves are
derived  for various learning algorithms.
The analytic results find excellent confirmation in  numerical
simulations.
\end{abstract}

\maketitle

\section{Introduction}
One of the more interesting properties of neural networks is their
ability to learn from examples.
In on-line learning processes a student network updates its
couplings after the presentation of each example in
order to make its outputs agree with the outputs of
the teacher. In the standard situation the student knows only the inputs and
the corresponding outputs of the teacher and has no further knowledge
of the rule used by the latter. Furthermore, in the course of learning the
student is able to classify correctly also new examples, which it has
never seen before. The latter property is called generalisation.

Various aspects of learning and generalisation in neural networks
have been intensively studied in many different contexts. For about a
decade now statistical mechanical methods have been used successfully in
these studies
(for recent reviews see, for example \cite{OK96,MC98,S98,EVdB01}).

A lot of the theoretical research has been concentrated on the
simplest models, such as the binary perceptron. Parallel to the progress
in these investigations, new more realistic models have been
considered, e.g., models with multi-state neurons \cite{WRBvM92}, models with
multi-neuron interactions \cite{BdAI95,YO98}, models with many layers 
(see, e.g, \cite{HMM92,K94,CC95}).

In this paper we study  on-line learning and generalisation in a
recently introduced  model, allowing two different types of binary
neurons at each site, possibly having different functions \cite{KB01,BK01}.
More specifically, this so called
Ashkin-Teller (AT) perceptron  contains, besides two-neuron interaction
terms, also a four-neuron interaction term.
For the underlying biological motivation for the
introduction of different types of neurons we refer to \cite{BK99}.
Here, we
recall that  the maximal capacity of the AT perceptron model II introduced
in \cite{KB01,BK01}
can be larger than the one of the standard binary perceptron  \cite{ BK01}
and that the corresponding recurrent network model can be a more efficient
associative memory than
a sum of two Hopfield models \cite{BK99}.
A natural question is then how this AT perceptron performs in on-line
learning and generalisation tasks.

Two learning scenarios turn out to be of interest.
In the first scenario where the student and the teacher are independent
AT perceptrons, we show that the resulting learning curves do not differ
very much from the already known ones for perceptrons with multi-state neurons.
For some particular values of the network
parameters we precisely reproduce the learning curve of the 4-state Potts
perceptron \cite{WRBvM92}.

In the second scenario both the student and the teacher are represented by
a simple perceptron but they are coupled by an AT type
four-neuron interaction term.
Hence, contrary to the standard setup, they are not independent. This
can be considered as a sort of ``hardware'' coupling. 
As a result, also the teacher mapping is changing in the process of
learning.
We obtain a set of learning curves which qualitatively
differ from those found in the independent setup. We also find different
asymptotic behaviour when the number of examples increases to infinity.
For certain values
of the network parameters such a coupling describes the
realistic situation that the rule used by the teacher is partially
shared by the student.

The rest of the paper is organised as follows. In section \ref{model}
the model and the learning scenarios are introduced. The  formulas for
the generalisation error are derived in section \ref{generr}. The
differential equations for the evolution of the order parameters are
obtained in section \ref{dif}. Their solutions, compared with numerical
simulations can be found in section \ref{results}. In section \ref{conclus}
some concluding remarks are presented. Finally, two appendices contain
some technical details of the derivations.

\section{The model and the learning scenarios}\label{model}
The AT perceptron is defined as a mapping of the binary ($\pm 1$) inputs
$\{s_i,\sigma_i\}$, $i=1,...,N$ into two binary ($\pm 1$) outputs
$s$ and $\sigma$:
\begin{eqnarray}
\fl s=\sgn (h_1)+\theta (\gamma_3|h_3|-\gamma_1|h_1|)
     \theta (\gamma_2|h_2|-\gamma_1|h_1|)(\sgn(h_2h_3)-\sgn(h_1))
\label{maps}\\
\fl \sigma=\sgn (h_2)+\theta (\gamma_3|h_3|-\gamma_2|h_2|)
     \theta (\gamma_1|h_1|-\gamma_2|h_2|)(\sgn(h_1h_3)-\sgn(h_2))~,
\label{mapsigma}
\end{eqnarray}
where $\theta$ is the Heaviside step function and $\gamma_r\geq 0$,
$r=1,2,3$, denote the
strength of the local fields $h_r$ which are defined as follows
\bea
&&  h_1 = \frac{1}{n_1} \sum_i J_i^{(1)} s_i,~~~~
    h_2= \frac{1}{n_2} \sum_i J_i^{(2)} \sigma_i,~~~~\nonumber\\
&&  h_3 = \frac{1}{n_3} \sum_i J_i^{(3)} s_i \sigma_i,~~~
    n_r^2 = \sum_i(J_i^{(r)})^2 \,. \label{fields}
\eea
The mapping (\ref{maps})-(\ref{mapsigma}) can be equivalently represented
by the set of three equations (cfr. model I in \cite{BK01})
\bea
  s&=&\sgn(\gamma_1h_1 +\sigma \gamma_3h_3 )\label{e.s1}\\
  \sigma&=&\sgn(\gamma_2h_2+s \gamma_3h_3)\label{e.s2}\\
  s\sigma&=&\sgn(\sigma \gamma_1h_1 +s \gamma_2h_2)\label{e.s3} \, .
\eea
For $\gamma_3=0$ the outputs $s$ and $\sigma$ are completely
independent and defined like in the simple perceptron
\bea
   s&=&\sgn (h_1)\label{smap0}\\
   \sigma&=&\sgn (h_2)\label{sigmamap0} \, .
\eea

\subsection{Learning scenario I}
First, we consider the standard situation where the student and the
teacher are two completely independent networks. In our case they are
represented by  AT perceptrons meaning that
the outputs of the teacher $\{s_T,\sigma_T\}$
and of the student $\{s_S,\sigma_S\}$ are both determined by the
mapping (\ref{maps})-(\ref{mapsigma}) but with  different
 couplings:
$\bJ_r^T$ and $\bJ_r^S$ respectively, with $\bJ_r=\{J_i^{(r)}\}$.
Initially, the student and the
teacher couplings are not correlated.  At each time step $t$ an example
is presented to the student. The student network then updates 
its couplings according to the following learning rule $F$
\begin{eqnarray}
  \bJ_1^S(t+1)&=&\bJ_1^S(t)+\frac{1}{N}Fs_T(t)\bs(t)\label{evj1}\\
  \bJ_2^S(t+1)&=&\bJ_2^S(t)+\frac{1}{N}F\sigma_T(t)\bsigma(t)\label{evj2}\\
  \bJ_3^S(t+1)&=&\bJ_3^S(t)+\frac{1}{N}Fs_T(t)\sigma_T(t)\bpsi(t)\label{evj3}
\end{eqnarray}
where
\bea
  \bs=\{s_i\},~~~\bsigma=\{\sigma_i\},~~~
  \bpsi=\{s_i\sigma_i\}~.
\eea
In this scenario we consider only Hebbian learning for which $F=1$.
Furthermore, examples are chosen randomly with equal probability out of
the complete set of examples.

\subsection{Learning scenario II}\label{scen2}
Alternatively, the AT perceptron can also be seen as two coupled
perceptrons, with outputs $s$ and
$\sigma$. In the second scenario we precisely analyse learning between such
coupled perceptrons (or branches of the AT perceptron). The outputs of the
student $s$ and the teacher $\sigma$ are defined by the equations
(\ref{maps}) and (\ref{mapsigma}) respectively. 

When $h_3>0$, the teacher and the student use two different mixtures of two
perceptron mappings defined by the couplings $\bJ_1$ and $\bJ_2$.
It implies that $s$ and $\sigma$ are always equal to $\sgn(h_1)$ or
$\sgn(h_2)$ and sometimes, depending on the relation between $\gamma_1h_1$,
$\gamma_2h_2$ and  $\gamma_3h_3$, $s=\sigma$.
In the limit $\gamma_3\rightarrow\infty$ the student and
the teacher network become so strongly coupled that one always has
$s=\sigma$ and the mapping (\ref{maps})-(\ref{mapsigma}) can be simplified to
\bea
  s=\sigma=\sgn (h) ~~~~~~~ h=\{h_x : |h_x|>|h_y|;
           x,y=1,2\} .\label{onerule+}
\eea
For $h_3<0$, the situation is quite different. Even with $\bJ_1=\bJ_2$ there
is always a non-zero fraction of disagreements between the student and
the teacher, as long as $\gamma_3>0$. In the limit $\gamma_3\to\infty$
the  student always disagrees with the teacher, and the mapping 
(\ref{maps})-(\ref{mapsigma}) can be written in the form:
\bea
s=\cases{
   -\sigma=\sgn(h_1)  ~~{\rm if} ~~|h_1|>|h_2|\cr
   -\sigma=-\sgn(h_2) ~~{\rm if}~~|h_1|<|h_2| } \, .
\label{onerule-}
\eea
For any value of the coupling field $h_3$ and $\gamma_3=0$ the student and
the  teacher are independent and they use the mappings defined by only one
coupling vector (cfr. (\ref{smap0})-(\ref{sigmamap0})).

In the sequel we take $\bs=\bsigma$ because the student
and the teacher must have the same inputs. We remark that this implies
that $h_3= \sum_i J_i^{(3)}/{n_3}$ (cfr.(\ref{fields})).
Again, at each time step $t$ an example is presented to the student network
and  its coupling vector $\bJ_1$ is updated as follows
\bea
    \bJ_1(t+1)=\bJ_1(t)+\frac{1}{N}F(\gamma_1h_1,\gamma_3h_3,s,\sigma)
    \sigma(t)\bs(t) \, .
\eea
Furthermore, at each time step a new coupling vector $\bJ_3$ is
generated
thus making the coupling between the perceptrons random. The coupling vector
of the teacher, $\bJ_2$, is not changed in the process of learning, but
later on we average over all possible teachers.
In this scenario we consider three learning rules $F$:
\begin{description}
\item \hspace*{2cm} Hebbian $F(\gamma_1h_1,\gamma_3h_3,s,\sigma)=1$
\item \hspace*{2cm} Perceptron $F(\gamma_1h_1,\gamma_3h_3,s,\sigma)=
\theta (-s\sigma)$
\item \hspace*{2cm} Adatron  $F(\gamma_1h_1,\gamma_3h_3,s,\sigma)=
-(\sigma \gamma_1h_1+\gamma_3h_3)
                                      \theta (-s\sigma)$
\end{description}

\section{Generalisation error}\label{generr}
A quantity of interest in the sequel is the generalisation error. It is
defined as the probability that the student and the teacher disagree,
i.e. that their outputs are different. When the teacher and the student
are simple independent perceptrons the
generalisation error $\varepsilon_g=\arccos (\rho)/\pi$ is a simple
function of the overlap $\rho=\bJ^T\cdot\bJ^S/(n^Sn^T)$ between the
student and the
teacher couplings, which in this case plays the role of an order parameter.
Unfortunately, for more complicated models this relation takes a much more
involved form (see, e.g., \cite{WRBvM92}).

\subsection{Scenario I}
In the first scenario the definition of the generalisation error reads
\be
   \varepsilon_g(\rho_1,\rho_2,\rho_3)=
        \left\langle1-\frac{1}{4}(1+s_Ts_S)(1+\sigma_T\sigma_S)
                 \right\rangle_I
\label{gerr},
\ee
with the overlaps $\rho_r$ defined by
\bea
   \rho_r=\frac{\bJ^S_r\cdot\bJ_r^T}{n^T_r n^S_r} \, ,
\eea
and with  $\langle \ldots \rangle_I=
\int\d{\bf h^T}\d{\bf h^S} \ldots P_I({\bf h^T},{\bf h^S})$
denoting the average over the teacher field, ${\bf
h^T}=\{h_1^T,h_2^T,h_3^T\}$, and the
student field, ${\bf h^S}=\{h_1^S,h_2^S,h_3^S\}$, which have a joint
probability distribution $P_I({\bf h^T},{\bf h^S}) $. The averages over these
fields are double averages, one over the examples and one over the
couplings. This arises because the couplings and the examples enter
the mapping (\ref{maps})-(\ref{mapsigma}) and the learning rules only
through the local fields.
We assume that the examples are taken randomly with equal probability
out of the full training set. Then, in the thermodynamic limit the local
fields become correlated Gaussian variables and the joint probability
distribution
$P_I({\bf h^T},{\bf h^S})$ can be written down in the form
\bea
   \fl P_I({\bf h^T},{\bf h^S})=
    \left((1-\rho_1^2)(1-\rho_2^2)(1-\rho_3^2)\right)^{-1/2}
   \frac{1}{2\pi^3}\exp\left\{
     \frac{\rho_1h_1^Sh_1^T}{1-\rho_1^2}+\frac{\rho_2h_2^Sh_2^T}
    {1-\rho_2^2}+\frac{\rho_3h_3^Sh_3^T}{1-\rho_3^2} \rb.\nn \\
       \lb. -\frac{1}{2}\left[\frac{(h_1^S)^2 + (h_1^T)^2}{1-\rho_1^2}+
   \frac{(h_2^S)^2+(h_2^T)^2}{1-\rho_2^2}+\frac{(h_3^S)^2+(h_3^T)^2}
           {1-\rho_3^2}\right]\right\}
    \label{p1}  \, .
\eea
Performing the averages in (\ref{gerr}) explicitly  leads to the expression
\bea
    \varepsilon_g(\rho_1,\rho_2,\rho_3)=\frac{3}{4}-\sum_{r=1}^3 I_r
\eea
with
\bea
\fl I_r=\frac{1}{2}\int_0^\infty \D h_r^T \erf\left(\frac{\rho_r
          h_r^T}{\sqrt{2(1-\rho_r^2)}}\right)\nn
       \left[1-2\left(1-\erf\left(\frac{\gamma_r h^T_r}
            {\gamma_{r'} \sqrt{2}}\right)\right)
     \left(1-\erf\left(\frac{\gamma_r h^T_r}{\gamma_{r''} \sqrt{2}}
                               \right)\right)\right]\nonumber
\\
  \fl \hspace*{0.5cm} +\frac{1}{4}\int \D(h_r^T,h_r^S) (a^+_{rr'}-a^-_{rr'})
           (a^+_{rr''}-a^-_{rr''})+
            (a^+_{rr'}+a^-_{rr'})(a^+_{rr''}+a^-_{rr''})
                             \sgn(h^T_r h^S_{r})\, , 
\\
\fl a_{rr'}^{\pm}= \frac{1}{2}\left(1-\erf\left(\frac{\gamma_r|h^T_r|}
      {\gamma_{r'}\sqrt{2}}\right)\right)
    -\int_{-\infty}^{-\frac{\gamma_r|h^T_r|}{\gamma_{r'}}}\D h^T_{r'}
      \erf\left(\frac{\gamma_r|h^S_r|\pm\gamma_{r'}\rho_{r'} h^T_{r'}}
        {\gamma_{r'} \sqrt{2(1-\rho_\nu^2)}}\right) \, ,
\eea
where $Dz= dz \exp(-z^2/2)/ \sqrt{2 \pi}$ is the Gaussian measure,
$r',r''=1,2,3$ ($r \neq r' \neq r'' \neq r$) and where
\bea
  \D(h_r^T,h_r^S)=\frac{\d h^T_r\d h^S_r}{2\pi\sqrt{1-\rho_r^2}}
         \exp\lb\{-\frac{1}{2}
     \frac{(h^T_r)^2+(h^S_r)^2-2\rho_r h^S_r h^T_r}{1-\rho_r^2}\rb\}
\eea
is a correlated Gaussian.

\subsection{Scenario II}
In the second scenario the generalisation error is given by
\bea
    \varepsilon_g(\rho)
    =\left\langle 1-\frac{1}{2}(1+s\sigma)\right\rangle_{II}
     =\int \d{\bf h}P_{II}({\bf h})\left(1-\frac{1}{2}(1+s\sigma)\right)
     \, ,
\eea
with the overlap $\rho$ defined by
\bea
  \rho=\frac{\bJ_1\cdot\bJ_2}{n_1 n_2} \, .
\eea 
Here again, as in the first scenario, the average over the examples and the
couplings is done through averaging over the local fields. The examples
are chosen randomly with equal probability out of the full set of
examples. In the thermodynamic limit this leads to a Gaussian
distribution of the local fields.
Since the behaviour of the system strongly depends on the sign of the
coupling field $h_3$ we consider three different field distributions $P_{II}$
\bea
P_\pm({\bf h})&=& {\left((2\pi)^3(1-\rho^2)\right)^{-1/2}}
\exp\left\{-\frac{1}{2}\left(
   \frac{h_1^2+h_2^2-2h_1h_2\rho}{(1-\rho^2}+h_3^2\right)\right\}
                 \label{ppm}\\
P_+({\bf h})&=&2\,\, P_\pm({\bf h})\,\, \theta (h_3)\label{pp}\\
P_-({\bf h})&=&2\,\, P_\pm({\bf h}) \,\,\theta (-h_3)\label{pm} \, .
\eea
In the case of the distribution $P_\pm$ the components of the vector
$\bJ_3$ are taken randomly (with equal probability) from some interval
$(-a,a)$,  with $a$ a positive real number.
In the case of the distributions $P_+$ and $P_-$ these components are
chosen in the same way but those values which lead to negative
respectively positive values of the field $h_3$ are omitted.
The generalisation error in these three situations reads, with obvious
notation
\be
   \varepsilon_g^c(\rho) =\frac{1}{\pi}\arccos (\rho)
                 +I_c~~~~~~~c=\pm,+,-\label{gerr2}
\ee
where
\bea
\fl I_\pm=\frac{1}{2}\left(u_{12}^--u_{12}^++u_{21}^--u_{21}^+\right)
           ,~~~I_+=-u_{12}^+-u_{21}^+,~~~I_-=u_{12}^-+u_{21}^- 
	   \label{Ies}               
\eea
and
\bea
  u_{rr'}^\pm&=&\int_{-\infty}^0\D h_2\left(1+\erf\left(
         \frac{\gamma_r h_2}{\gamma_3 \sqrt{2}}\right)\right)
         \left(1+\erf\left(\frac{h_2\left(\frac{\gamma_r}{\gamma_r'}\pm\rho
     \right)}
            {\sqrt{2\left(1-\rho^2\right)}}\right)\right)\, .
        \label{integ2}
\eea
It is easy to realize that only for positive $h_3$ (i.e. for $P_{II}=P_+$)
the generalisation error $\varepsilon_g^+(\rho)$ goes to zero as $\rho$ 
goes to 1. It is also equal to zero for any $\rho$ when $P_{II}=P_+$ and 
$\gamma_3=\infty$.

\section{Order parameters and their evolution}\label{dif}
As can be seen from the formulas written down in the last section, the
generalisation error is a function of the overlaps $\rho$ or $\rho_r$,
which play the role of order parameters in the learning process. Their
evolution is coupled with the evolution of the norms of the couplings $n_r$
and in the thermodynamic limit $N \rightarrow \infty$ it can be
described  by ordinary differential equations \cite{KC92}.

In the first scenario  a standard calculation (for a review see, e.g., 
\cite{MC98}) leads to the following result for Hebbian learning   
\bea
 \fl \frac{\d}{\d \alpha}n_r=\left\langle\Psi^T_r h_r^S\right\rangle_I
                 +\frac{1}{2 n_r}~~~~~
\frac{\d}{\d \alpha}\rho_r=\frac{1}{n_r}\left\langle\Psi^T_r(h_r^T-
         \rho_r h_r^S )\right\rangle_I-\frac{\rho_r}{2n_r^2}
               ~~~r=1,2,3
               \label{dif1}
\eea
where $\Psi^T_1=s_T$, $\Psi^T_2=\sigma_T$, $\Psi^T_3=s_T\sigma_T$ and
$\alpha=t/N$ is the number of examples scaled with the size of the
system. It becomes continuous in the thermodynamic limit.
After performing the averages we arrive at
\bea
 \frac{\d n_r}{\d \alpha}=\rho_r b_r+\frac{1}{2n_r}~~~~~
 \frac{\d \rho_r}{\d \alpha}=\frac{1-\rho^2}{n_r}b_r-\frac{\rho_r}{2n_r^2}
             \label{dif1a}
\eea
with the quantity $b_r$ given by
\bea
 b_r&=&\sqrt{\fr{2}{\pi}}\lb\{\fr{1}{\sqrt{c_{r'r}}}
          \lb[1-2\int_0^\infty\D h~\erf\lb(\fr{h\gamma_{r'}}
           {\gamma_{r''}\sqrt{2c_{r'r}}}\rb)\rb]  \rb.\nn
\\
   &+&\lb. \fr{1}{\sqrt{c_{r''r}}}
        \lb[1-2\int_0^\infty\D h~\erf\lb(\fr{h\gamma_{r''}}
            {\gamma_{r'}\sqrt{2c_{r''r}}}\rb)\rb]\rb\}
\\
        c_{rr'}&=&1+\lb(\fr{\gamma_r}{\gamma_{r'}}\rb)^2~.
	\label{defc}
\eea
For $\gamma_1=\gamma_2=\gamma_3$ this quantity simplifies to
\bea
     b_r=2\lb(1-\frac{2}{\pi}\arctan\lb(\frac{1}{\sqrt{2}}\rb)\rb)
               \cong 1.21635 \, .
\eea
We remark that the differential equations (\ref{dif1a}) for a given $r$
have the same form as the differential equations found for the simple
perceptron with Hebbian learning \cite{MC98}. More specifically, they
differ only by the value of the coefficient $b_r$,  which for the
simple perceptron is equal to $\sqrt{{2}/{\pi}}\cong 0.798$.

For the Hebbian learning we are considering, it is possible to construct a
simple expression for $\rho_r$ as a function of $\alpha$. Following Opper
and Kinzel \cite{OK96} we
slightly modify the update rule (\ref{evj1}), (\ref{evj2}), (\ref{evj3})
(substituting $1/N$ by $1/\sqrt{N}$) and easily arrive at
\be
  \rho_r=\sqrt{\frac{\alpha a_r^2}{\alpha a_r^2 +\pi}}\label{easyro}
\ee
where we have taken as initial condition $\rho (0)=0$ and where
\bea
\fl a_r=2\sqrt{\pi}\int_0^\infty\D b~b\left[1-
\left(1-\erf\left(\frac{\gamma_r b }{\gamma_{r'}\sqrt{2}}\right)\right)
\left(1-\erf\left(\frac{\gamma_r b }{\gamma_{r''}\sqrt{2}}\right)\right)
\right]~.
\eea
This expression differs from the solution of (\ref{dif1a}) only for small
values of $\alpha$ and has the advantage of having a simple form.
The evolution of $\rho$ in the case of simple perceptrons is described by
the single equation (\ref{easyro}), but with a coefficient $a_r=\sqrt{2}$.
Since these results are very similar to the results obtained for the simple
perceptron we do not test other algorithms in this scenario because we
expect that also in those cases  a strong resemblance to the simple 
perceptron occurs.

In the second scenario with the learning rule F defined in  subsection
\ref{scen2} we have to solve the following set of differential equations:
\bea
   \fl\frac{\d}{\d\alpha}n_1&=&\langle h_1\sigma 
   F(\gamma_1h_1,\gamma_3h_3,s,\sigma)\rangle_{II}
     +\frac{1}{2n_1}\left\langle 
     F^2(\gamma_1h_1,\gamma_3h_3,s,\sigma)\right\rangle_{II}\label{dif2a}
     \\
\fl   \frac{\d}{\d\alpha}\rho&=&\frac{1}{n_1}\left\langle\sigma 
   F(\gamma_1h_1,\gamma_3h_3,s,\sigma)
       \left(h_2-\rho h_1\right)\right\rangle_{II}-\frac{\rho}{2 n_1^2}
                           \left\langle 
		F^2(\gamma_1h_1,\gamma_3h_3,s,\sigma)\right\rangle_{II}.
             \label{dif2b}
\eea
Performing the averages leads to much more complicated expressions
 than those obtained in the first scenario.
The explicit form of these expressions obtained for Hebbian, perceptron
and Adatron learning with the  distributions $P_\pm$ and $P_+$
can be found in \ref{explicit}.

\section{Results}\label{results}
In this section we discuss the numerical solutions of the differential
equations (\ref{dif1}) and (\ref{dif2a})-(\ref{dif2b}) and compare them 
with the results of
simulations. Because only the ratios of the strength
parameters $\gamma_1$, $\gamma_2$ and $\gamma_3$ are important
we take $\gamma_1=\gamma_2=1$, and vary only $\gamma_3$.

\subsection{Scenario I}
The learning curves for small values of the number of examples $\alpha$
obtained in the first scenario using formula (\ref{easyro}) are presented
in figure \ref{atp2atp}. All curves start with an initial generalisation
error $\varepsilon_g=0.75$ corresponding to random guessing in four-state
models.
For $\gamma_3=0$ learning between two  independent perceptrons is
described. For $\gamma_3=1$ the learning curve is identical with the
one of the 4-state Potts perceptron \cite{WRBvM92} (cfr. \cite{KB01,BK01}).
In the limit $\alpha\rightarrow\infty$, $\varepsilon_g$ decays like 
$\alpha^{-\frac{1}{2}}$ for all values of $\gamma_3$, precisely like in the
case of learning between simple perceptrons.

\subsection{Scenario II}
A careful analysis of expression (\ref{gerr2}) leads to the conclusion that
in the second scenario the generalisation error can be nonzero even when the
normalised angle between the student and the teacher couplings, $\phi=\arccos
(\rho)/\pi$, is equal to zero. This happens when we allow the field $h_3$
to take  negative values. Therefore, we follow the evolution of two dynamical
variables in the sequel: the generalisation error $\varepsilon_g$ and
the normalised angle between the student and the teacher $\phi$.
For all the learning algorithms and distributions of the fields 
that we have considered, we observe an abrupt change in the asymptotic
behaviour in $\alpha$ when $\gamma_3$ changes from $0$ to some non-zero
value. 
Logarithmic plots of the learning curves for two distributions of the
fields, $P_\pm$ and $P_+$, are presented in figures 
\ref{batp2batppm1}-\ref{batp2batpp3}. The learning curves for the
distribution $P_-$ are qualitatively very similar to the curves obtained
for $P_\pm$.

\subsubsection{$P_{II}=P_\pm$}
Let us first analyse the results obtained for the distribution $P_\pm$
in more detail. For $\gamma_3\not= 0$, the generalisation error
saturates at some non-zero
value. For Hebbian and perceptron learning the angle $\phi$ between the
student and the teacher is asymptotically decreasing to zero
at a higher rate than in the decoupled case $\gamma_3=0$. For Hebbian  
learning we find that in the limit $\alpha\rightarrow\infty$, 
 $\phi\sim\alpha^{-1}$, versus $\phi\sim\alpha^{-\frac{1}{2}}$ for 
 $\gamma_3=0$, while in the case of the perceptron algorithm 
$\phi\sim\alpha^{-\frac{1}{2}}$, versus  $\phi\sim\alpha^{-\frac{1}{3}}$ 
for $\gamma_3=0$. For the Adatron algorithm  $\phi$ and 
$\varepsilon_g$ both saturate at some non-zero value.
In spite of the fact that the generalisation error never vanishes
the student is able to learn the couplings of the teacher
using the Hebbian or perceptron algorithm.

\subsubsection{$P_{II}=P_+$}
We observe that for all algorithms the generalisation error goes
asymptotically  to zero. For Hebbian and perceptron learning it 
decreases  faster than in the decoupled case. In the limit 
$\alpha\rightarrow\infty$, we get $\varepsilon_g\sim\alpha^{-1}$ for
Hebbian learning  while $\varepsilon_g\sim\alpha^{-\frac{1}{2}}$ for
perceptron  learning. For Adatron learning
we obtain the same decay exponent as in the decoupled case.
Surprisingly, for the perceptron and Adatron algorithms  the decay of the
angle between the student and the teacher, $\phi$, is slower than in
the decoupled case in the limit  $\alpha\rightarrow\infty$. For the
perceptron  we have
$\phi\sim\alpha^{-\frac{1}{4}}$ and for the Adatron we find
$\phi\sim\alpha^{-\frac{1}{2}}$. On the contrary, for Hebbian learning 
$\phi\sim\alpha^{-\frac{1}{2}}$ as for the decoupled case.

Since an analytic analysis of the differential equations (see
\ref{explicit}) is rather involved, the asymptotic exponents
discussed above have been determined numerically.    
Only in the case of Hebbian learning with the field distribution
$P_+$ the numerical analysis was not entirely unambiguous. Therefore, we have
derived the corresponding exponents analytically. Details 
can be found in \ref{asympt}.

The initial generalisation error is a function of the strength parameter
$\gamma_3$, which measures the strength of the coupling between the two
perceptrons.  The larger the $\gamma_3$ the bigger the common knowledge
between  the student and the teacher, so the smaller the initial error. For 
$\gamma_3\rightarrow\infty$, the
student and the teacher use precisely the same rule (\ref{onerule+})
in order to determine their outputs.

Finally, the numerical solution of the equations (\ref{dif2a})-(\ref{dif2b})
suggests that there is a 
simple relation between the decay exponents of $\phi$  and
$\varepsilon_g$, denoted by $y_\phi$ and $y_g$ respectively,
\be
   y_g=2y_\phi\label{exprel} \, .
\ee
This relation can also be derived analytically (see \ref{asympt}).
For $\gamma_1=\gamma_2$ we find in the limit $\alpha\rightarrow\infty$ 
(and $\phi\rightarrow 0$) that 
\bea
\varepsilon_g^+\sim\frac{\pi^2}{4\sqrt{2}\gamma_3}\phi^2~,
\eea
confirming the observation (\ref{exprel}).

\subsection{Computer simulations}
To check the analytic results described above we have performed numerical 
simulations. The system sizes have been varied between $N=100$ and
$N=999$ neurons. 
An excellent agreement has been found for both scenarios and all learning 
algorithms, even for relatively small $N$.  
As a representative example we present a comparison between simulations
and analytic results obtained in the second scenario with the Adatron
algorithm  for $\gamma_3=0.1$ and $P_{II}=P_{\pm}$.
For the sake of clarity we show the results obtained for small  and 
big $\alpha$ separately. The analytic results for small $\alpha$
are compared with simulations for a system with $N=999$ neurons 
(fig. \ref{simad2}). For bigger $\alpha$ we have made simulations for smaller 
systems ($N=100$), which are displayed in fig. \ref{simad2bis}.
In both cases only the results obtained for one sample are shown.

For small $\alpha$ the simulations are smoothly aligned along the
theoretical curves. This points to the self averaging property of the
learning process. For bigger values of $\alpha$ very strong fluctuations
occur around the theoretical result. This happens only for the Adatron  
algorithm and $P_{II}=P_{\pm}$ and, hence, cannot be explained entirely
by  the relatively small size of the system.  Indeed, as has been noticed in
section \ref{results}, in this case  there is always a non-zero fraction of
disagreement between the student and the teacher. So, a strategy used by the
Adatron algorithm which updates the couplings proportional to the error
made by the student, must lead to rather big random changes. Nevertheless the
simulation points in fig. \ref{simad2bis} are evenly distributed on both
sides of the theoretical curve.

\section{Conclusions}\label{conclus}
In this paper we have studied on-line learning and generalisation using
the AT perceptron. Two learning scenarios have been considered.
The results obtained in the first scenario, where the student and the
teacher are represented by independent AT perceptrons, are very similar
to the results obtained for the simpler models \cite{MC98}. For a
particular choice of the network parameters the learning curve precisely 
reproduces that found for the 4-state Potts perceptron \cite{WRBvM92}.

In the second scenario the student and the teacher are taken to be simple 
perceptrons coupled by a four-neuron interaction term. 
Particular results depend crucially on the
distribution of the couplings $\bJ_3$.

For the field distribution $P_{II}=P_\pm$ the generalisation error always
saturates at some non-zero value. This is not surprising since this
distribution allows the field $h_3$ to take negative values what inevitably
leads to a non-vanishing fraction of disagreements between the student
and the teacher even when $\bJ_1=\bJ_2$ (cfr. (\ref{maps}), 
(\ref{mapsigma})). In spite of this, for Hebbian and perceptron learning
the student manages to learn the couplings of the teacher perfectly (in the 
limit $\alpha\rightarrow\infty$). This does not happen, however, for the 
Adatron algorithm, which in the standard (decoupled) situation proved to be 
the fastest \cite{MC98}. The reason is that this algorithm changes the
couplings of the student proportionally to the error made by the latter. 
Since this error is non-zero even for $\bJ_1=\bJ_2$, this cannot be a good
strategy. Hence, the more "blind" updates (Hebbian and perceptron)
appear to be more effective.

For $P_{II}=P_+$ we have obtained quite different results. In this case the 
generalisation error goes to zero when $\rho$ goes to $1$. For Hebbian and 
perceptron learning we observe faster decay of $\varepsilon_g$ than in the 
decoupled case. For Adatron learning the decay exponent of $\varepsilon_g$ 
is the same as for $\gamma_3=0$. Surprisingly, for all algorithms we find 
the same or slower decay of $\phi$ compared with the decoupled case.

The best asymptotic decay of the generalisation error has been obtained for
$P_{II}=P_+$ with the Adatron rule: $\varepsilon_g\sim 0.618\alpha^{-1}$.
Comparing  with the case of independent perceptrons we see that it is
better than the lower bound for on-line learning \cite{MC98}
($\varepsilon_g\sim 0.88\alpha^{-1}$) and worse than the Bayesian 
lower bound \cite{OH91} ($\varepsilon_g\sim 0.44\alpha^{-1}$).

We remark that in the course of a learning process in the second scenario also
the teacher mapping is changed but not the teacher couplings. This can
be interpreted as a kind of effective mutual learning caused by the
(``hardware'') coupling  of the two perceptrons. This is different from the 
mutual learning process analysed in \cite{KMK00,MKK00}, the only other
learning process of this type  known to us. There, in contrast to our setup, 
the teacher explicitly  learns from the student.
In our model the decay exponent of $\varepsilon_g$ is not influenced by a
particular value of the strength parameter $\gamma_3$ as long as it is
nonzero.

The model analysed in the second scenario with $P_{II}=P_+$  where a part of
the learning rule is shared by
the teacher and the student, can be compared to a real life situation in
which both of them, e.g., have the same cultural background, followed the same 
education $\ldots$ 
One can expect that in such a situation the learning process is much
more efficient since the  student and the teacher speak in a sense the
same language. It corresponds to a faster asymptotic decay of the
generalisation error in our model. It
would be interesting to see, e.g.,  whether an optimisation of the learning
process \cite{KC92}  would still improve these results.  

\ack
This work was supported in part by the Fund for Scientific Research,
Flanders (Belgium).\\

\appendix
\section{The evolution of the order parameters in the second learning 
scenario}\label{explicit}
The set of differential equations (\ref{dif2a})-(\ref{dif2b}) for the order 
parameters
in the second learning scenario can  be written down in the following form:
\bea
\frac{\d n_1}{\d\alpha}&=&f_1(\rho,\gamma_{13},
\gamma_{23})+\frac{1}{2n_1}
f_2(\rho,\gamma_{13},
\gamma_{23})\nn
\\
\frac{\d\rho}{\d\alpha}&=&\frac{1}{n_1}
f_3(\rho,\gamma_{13},
\gamma_{23})-\frac{\rho}{2n_1^2}
f_2(\rho,\gamma_{13},
\gamma_{23})\nn~,
\eea
with $\gamma_{rr'}=\gamma_r/\gamma_{r'}$ and
where the explicit form of $f_1(\rho,\gamma_{13},\gamma_{23})$, 
$f_2(\rho,\gamma_{13},\gamma_{23})$ and 
$f_3(\rho,\gamma_{13},\gamma_{23})$ depends on the 
algorithm used and on the distribution of the fields.

In the case of the distribution  $P_{II}=P_\pm$ we have for 
\begin{description}
\item Hebbian learning 
\bea
f_1(\rho,\gamma_{13},\gamma_{23})&=&\rho f_{21}+g_{21}\nn
\\
f_2(\rho,\gamma_{13},\gamma_{23})&=&1\nn
\\
f_3(\rho,\gamma_{13},\gamma_{23})&=&f_{21}(1-\rho^2)-\rho g_{21}\nn
\eea
\item Perceptron learning
\bea
f_1(\rho,\gamma_{13},\gamma_{23})&=&\frac{1}{2}(\rho f_{21}-f_{12}+g_{21})\nn
\\
f_2(\rho,\gamma_{13},\gamma_{23})&=&\frac{1}{\pi}\arccos (\rho)+I_\pm\nn
\\
f_3(\rho,\gamma_{13},\gamma_{23})&=&\frac{1}{2}
(f_{21}(1-\rho^2)-g_{12}-\rho g_{21})\nn
\eea
\item Adatron learning
\bea
\fl f_1(\rho,\gamma_{13},\gamma_{23})&=&
-\gamma_1 \left(f_a-f_{12}^++f_{12}^-+\frac{1}{2}\right)\nn
\\
\fl &-&\gamma_3 \left(t_{12}-\rho t_{21}+\frac{\sqrt{1-\rho^2}}{2\pi}\left(
\frac{1}{\sqrt{c_{21}^-}}+\frac{1}{\sqrt{c_{21}^+}}\right)\right)\nn
\\
\fl f_2(\rho,\gamma_{13},\gamma_{23})&=&
\gamma_1^2\left(f_a-f_{12}^++f_{12}^-+\frac{1}{2}\right)
-\gamma_3^2\left(\frac{1}{\pi}\arcsin
(\rho)-I_\pm-\frac{1}{2}\right)\nn
\\
\fl &+&\gamma_1\gamma_3\left( t_{12}-2\rho t_{21}+\frac{\sqrt{1-\rho^2}}{\pi}
\left(\frac{1}{\sqrt{c_{21}^-}}+\frac{1}{\sqrt{c_{21}^+}}\right)\right)
-\gamma_2\gamma_3 t_{21}\nn
\\
\fl f_3(\rho,\gamma_{13},\gamma_{23})&=&
\gamma_1\left(\frac{1}{\pi}\left(\sqrt{1-\rho^2}+\rho\arcsin 
(\rho)\right)+g^a_{21}+g^a_{12}+\rho\left(f_a+f_{21}^+-f_{21}^-
\right)\right)\nn
\\
&+&\gamma_3\left( t_{21}(1-\rho^2) + \frac{\sqrt{1-\rho^2}}{2\pi}
\left(\frac{1}{\sqrt{c_{12}^-}}+\frac{1}{\sqrt{c_{12}^+}}+
\frac{\rho}{\sqrt{c_{21}^-}}+\frac{\rho}{\sqrt{c_{21}^+}}\right)\right)\nn
\eea
\end{description}
with
\bea
\fl f_{rr'}=\sqrt{\frac{2}{\pi}}-\int_0^\infty\D h_r ~h_r
\left(1-\erf\left(\frac{\gamma_{r3} h_r}{\sqrt{2}}\right)\right)\nn
\\
\times\left[2-\erf\left(\frac{h_r\left(\gamma_{rr'}+\rho\right)}
{\sqrt{2(1-\rho^2)}}\right)
-\erf\left(\frac{h_r\left(\gamma_{rr'}-\rho\right)}
{\sqrt{2(1-\rho^2)}}\right)\right]\nn~,
\\
\fl g_{rr'}=\frac{1}{b_{rr'}}\sqrt{\frac{1-\rho^2}{2\pi}}
\left(1-\frac{2}{\pi}\arctan \left(\frac{\gamma_{r3}}{b_{rr'}}
\right)\right)
-\frac{1}{a_{rr'}}\sqrt{\frac{1-\rho^2}{2\pi}}
\left(1-\frac{2}{\pi}\arctan \left(\frac{\gamma_{r3}}{a_{rr'}}
\right)\right)\nn~,
\\
\fl g^a_{rr'}=\frac{\sqrt{1-\rho^2}}{2\pi}\left\{
\frac{1}{a_{rr'}^2}\left[\left(1+\gamma_{3r}^2
a_{rr'}^2\right)^{-\frac{1}{2}}-1\right]
+\frac{1}{b_{rr'}^2}\left[\left(1+\gamma_{3r}^2
b_{rr'}^2\right)^{-\frac{1}{2}}-1\right]\right\}\nn~,
\\
\fl f_a=\int_{-\infty}^0\D h_1 ~h_1^2
\erf\left(\frac{h_1\rho}{\sqrt{2(1-\rho^2)}}\right)\nn
+2\left(1-\rho^2\right)^{\frac{3}{2}}
\\
\times\int_0^\infty\D h_2 \left(1-\erf\left(\frac{\gamma_{23}
h_2}{\sqrt{2}}\sqrt{1-\rho^2}\right)\right)
\int_{\gamma_{21} h_2}^\infty\D h_1 ~h_1^2 \sinh (\rho h_1
h_2)\nn~,
\\
f_{rr'}^\pm=\frac{1}{2}\int_{-\infty}^0\D h_r~ h_r^2
\left(1+\erf\left(\frac{\gamma_{r3} h_r}{\sqrt{2}}\right)\right)
\erf\left(\frac{h_r\left(\gamma_{rr'}\pm\rho\right)}
{\sqrt{2(1-\rho^2)}}\right)\nn~,
\\
t_{rr'}^\pm=-\frac{1}{2\pi c_{r 3}}\left(\frac{c_{r 3}(1-\rho^2)}
{\left(\gamma_{rr'}\pm\rho\right)^2}+1\right)^{-\frac{1}{2}}
+\frac{1}{c_{r 3} 2\pi}~, \quad t_{rr'}=t_{rr'}^+-t_{rr'}^-\nn~,
\\
c_{r r'}^\pm=1+\left(\gamma_{r3}\right)^2
+\frac{\left(\gamma_{rr'}\pm\rho\right)^2}
{1-\rho^2}~,~~~
a_{rr'}=\sqrt{\frac{1+\left(\gamma_{rr'}\right)^2
-2\gamma_{rr'}\rho}{1-\rho^2}}\nn~,
\\
b_{rr'}=\sqrt{\frac{1+\left(\gamma_{rr'}\right)^2
+2\gamma_{rr'}\rho}{1-\rho^2}}\nn~,
\eea
where $I_{\pm}$ is given by expression (\ref{Ies}) and $c_{r3}$ is
defined in expression (\ref{defc}).

In the case of the distribution  $P_{II}=P_+$ we have for
\begin{description}
\item Hebbian learning
\bea
f_1(\rho,\gamma_{13},\gamma_{23})&=&\rho f_{21}^{'+}+g_{21}^+\nn
\\
f_2(\rho,\gamma_{13},\gamma_{23})&=&1\nn
\\
f_3(\rho,\gamma_{13},\gamma_{23})&=&(1-\rho^2)f_{21}^{'+}-\rho g_{21}^+\nn
\eea
\item Perceptron learning
\bea
f_1(\rho,\gamma_{13},\gamma_{23})&=&
\frac{1}{2}\left(\rho f_{21}^{'+}-f_{12}^{'+}+g_{21}^+\right)\nn
\\
f_2(\rho,\gamma_{13},\gamma_{23})&=&
\frac{1}{\pi}\arccos (\rho) + I_+\nn
\\
f_3(\rho,\gamma_{13},\gamma_{23})&=&
\frac{1}{2}\left(f_{21}^{'+}(1-\rho^2)-g_{12}^+-\rho g_{21}^+\right)\nn
\eea
\item Adatron learning
\bea
\fl f_1(\rho,\gamma_{13},\gamma_{23})&=&
-\gamma_1\left[f_a^+-2f_{12}^+-g_a+\frac{1}{2}\right]\nn
\\
\fl&+&\gamma_3\left[\frac{1}{\pi}(1-\rho)-2t_{12}^++2\rho t_{21}^+
-\frac{1}{\pi}
\sqrt{\frac{1-\rho^2}{c_{21}^+}}\right]\nn
\\
\fl f_2(\rho,\gamma_{13},\gamma_{23})&=&
\gamma_1^2\left[f_a^+-2f_{12}^+-g_a+\frac{1}{2}\right]
-\gamma_3^2\left[\frac{1}{\pi}\arcsin
(\rho) - I_+ -\frac{1}{2}\right]\nn
\\
&+&\gamma_1\gamma_3\left[-\frac{2}{\pi}(1-\rho)+2t_{12}^+-4\rho
t_{21}^++\frac{2}{\pi}\sqrt{\frac{1-\rho^2}{c_{21}^+}}
\right]-2\gamma_2\gamma_3t_{21}^+\nn
\\
\fl f_3(\rho,\gamma_{13},\gamma_{23})&=&
\gamma_1\left[\frac{1}{\pi}\left(\sqrt{1-\rho^2}+\rho\arcsin 
   (\rho)\right)+\rho f_a^++\rho g_a+2\left(g^b_{12}+g^b_{21}+
\rho f_{21}^+\right)\right]\nn
\\
&+&\gamma_3\left[2t_{21}^+(1-\rho^2)-\frac{1}{\pi}\left(1-\rho^2-
\sqrt{\frac{1-\rho^2}{c_{12}^+}}-\rho\sqrt{\frac{1-\rho^2}{c_{21}^+}}
\right)\right]\nn~,
\eea
\end{description}
with
\bea
\fl f_a^+=\int_{-\infty}^0\D h_1 ~h_1^2\erf\left(\frac{\rho h_1}
{\sqrt{2(1-\rho^2)}}\right)
-2\left(1-\rho^2\right)^{\frac{3}{2}}\nn
\\
\times\int_{-\infty}^0\D h_2\left(1+
\erf\left(\frac{\gamma_{23} h_2}{\sqrt{2}}\sqrt{1-\rho^2}
\right)\right)
\int_{-\infty}^{-\gamma_{21}|h_2|}\D h_1~ h_1^2
\exp \left(-\rho h_1 h_2\right)\nn~,
\\
\fl f_{rr'}^{'+}=\sqrt{\fr{2}{\pi}}+2\int_{-\infty}^0\D h~h
\lb(1+\erf\lb(\fr{\gamma_{r3} h}{\sqrt{2}}\rb)\rb)
\lb[1+\erf\lb(\fr{h\lb(\gamma_{rr'}+\rho\rb)}
{\sqrt{2(1-\rho^2)}}\rb)\rb]\nn~,
\\
\fl g_a=\int_{-\infty}^0\D h_1~ h_1^2\left[1+\erf\left(\frac{\gamma_{13} h_1}
{\sqrt{2}}\right)\right]\nn
~,~~
g_{rr'}^+=\frac{2}{b_{rr'}}\sqrt{\frac{1-\rho^2}{2\pi}}
\left(1-\frac{2}{\pi}\arctan\left(\frac{\gamma_{r3}}{b_{rr'}}\right)\right)
\\
g^b_{rr'}=\frac{\sqrt{1-\rho^2}}{2\pi}\frac{1}{b_{rr'}^2}
\left[\left(1+\gamma_{3r}^2b_{rr'}^2
\right)^{-\frac{1}{2}}-1\right]\nn \, .
\eea
and where $I_+$ is given by expression (\ref{Ies}).
\section{The asymptotic form of the solution in the
second scenario for Hebbian learning with $P_{II}=P_+$}\label{asympt}

Because the dependence of the generalisation error $\varepsilon_g^+$ on the
overlap $\rho$ is rather complicated (see (\ref{gerr2})) we derive the  
asymptotic form for $\varepsilon_g^+$ in two steps. First, we find the  
asymptotic
relation between $\varepsilon_g^+$ and $\phi$ and then we determine the
behaviour of $\phi$ as a function of $\alpha$ in the limit $\alpha\to\infty$.

\subsection{Asymptotic relation between $\varepsilon_g^+$ and $\phi$}
The generalisation error $\varepsilon_g^+$ is defined as (see (\ref{gerr2})):
\bea
\varepsilon_g^+=\fr{1}{\pi}\arccos\rho-u_{12}^+-u_{21}^+=
\phi-u_{12}^+-u_{21}^+\nn \, ,
\eea
with the integrals $u_{12}^+$, $u_{21}^+$ given by (\ref{integ2}).
We now expand these integrals as a function of $\phi$ for small
values of $\phi$. First, we change the variables to get
\bea
\fl u_{rr'}^+=\phi\int_{-\infty}^0\fr{e^{-\fr{1}{2}x^2\phi^2}}{\sqrt{2\pi}}
\d x (1+\erf(a\phi x))(1+\erf(cx))
 \equiv \phi\int_{-\infty}^0\d x f(\phi ,x)\nn~,
\eea
where
\bea
a=\fr{\gamma_r}{\gamma_3\sqrt{2}}~,~~~
c=\fr{\gamma_{rr'}+1}{\pi\sqrt{2}}~.\nn
\eea
Expanding $f(\phi,x)$ with respect to $\phi$ and taking
$\gamma_1=\gamma_2=1$ we get
\bea
u_{rr'}^+=\phi\fr{1}{\sqrt{2}\pi c}-\phi^2\fr{\sqrt{2}a}{4c^2\pi}-o(\phi^3)\nn
\eea
and this leads to
\be
\varepsilon_g^+=\fr{\pi^2}{4\sqrt{2}\gamma_3}\phi^2+o(\phi^3)\label{epsfi}~.
\ee

\subsection{Asymptotic relation between $\phi$ and $\alpha$}

The differential equations (\ref{dif2a})-(\ref{dif2b}) 
can be written down in terms of the
variables $n_1$ and $\phi$.
For Hebbian learning and $P_{II}=P_+$ this gives
\bea
\fr{\d n_1}{\d \alpha}&=&f_1(\cos(\pi\phi),\gamma_{13},\gamma_{23})
+\fr{1}{2n_1}f_2(\cos(\pi\phi),\gamma_{13},\gamma_{23})\label{difa1}
\\
\fr{\d \phi}{\d \alpha}&=&-\fr{f_3(\cos(\pi\phi),\gamma_{13},\gamma_{23})}
{n_1\pi\sin(\pi\phi)}
+\fr{\cos(\pi\phi)}{2\pi n_1^2\sin(\pi\phi)}f_2(\cos(\pi\phi),
\gamma_{13},\gamma_{23})\label{difa2}
\eea
The functions $f_1(\cos(\pi\phi),\gamma_{13},\gamma_{23})$, 
$f_2(\cos(\pi\phi),\gamma_{13},\gamma_{23})$ and 
$f_3(\cos(\pi\phi),\gamma_{13},\gamma_{23})$ are defined in \ref{explicit}.
Expanding the r.h.s of the differential equations 
(\ref{difa1})-(\ref{difa2}) around $\phi=0$ up to the
first non-vanishing term we can easily find that for $\gamma_1=\gamma_2$ 
\bea
\phi=\sqrt{\fr{\sqrt{2}}{2\pi(\sqrt{2}-1)}}\alpha^{-\fr{1}{2}}\nn
\eea
Combining this result with (\ref{epsfi}) we obtain the asymptotic formula for
the generalisation error:
\bea
\varepsilon_g^+=\fr{\pi}{8\gamma_3(\sqrt{2}-1)}\alpha^{-1}+
                          o(\alpha^{-\fr{3}{2}})\nn
\eea

\newpage

\renewcommand{\thefigure}{\arabic{figure}}

\begin{figure}[h]
\epsfysize=7cm
\epsfxsize=10cm
\centerline{\epsfbox{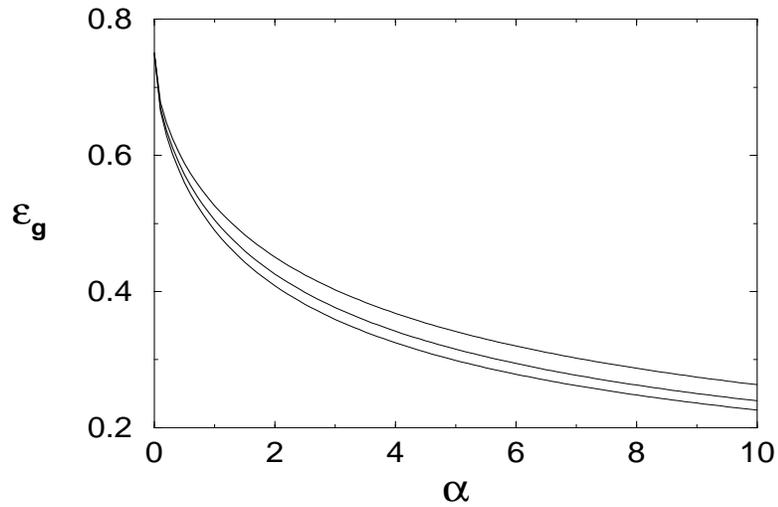}}
\caption{Learning scenario I: the generalisation error $\varepsilon_g$ 
as a function of the number of examples
$\alpha$ with $\gamma_1=\gamma_2=1$ and $\gamma_3=\infty, 1, 0$ from top
to bottom}
\label{atp2atp}
\end{figure}

\begin{figure}[h]
\epsfysize=8cm
\epsfxsize=10cm
\centerline{\epsfbox{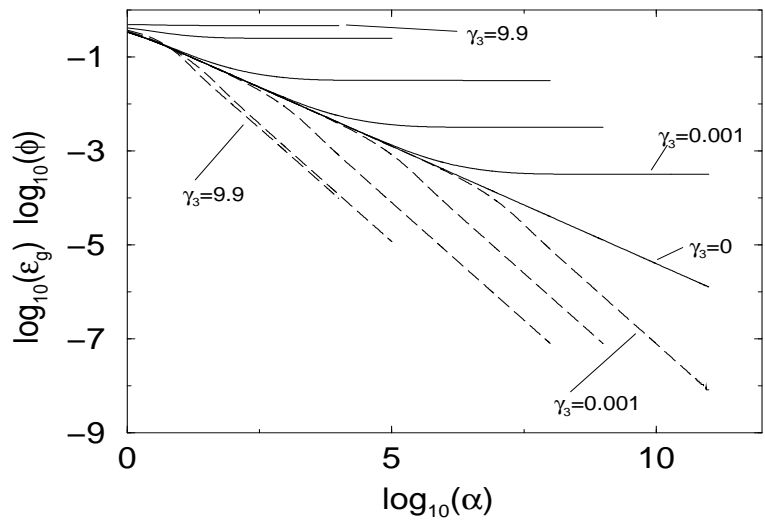}}
\caption{Learning scenario II: Log-log plot of the generalisation error, 
$\varepsilon_g$ (solid lines), and the normalized angle between the teacher 
and the student, $\phi$ (broken line), for $P=P_\pm$ and Hebbian
learning as a function of the number of examples $\alpha$.
Intermediate curves not marked on the figure are for $\gamma_3=0.01,0.1,1$}
\label{batp2batppm1}
\end{figure}

\begin{figure}[h]
\epsfysize=8cm
\epsfxsize=10cm
\centerline{\epsfbox{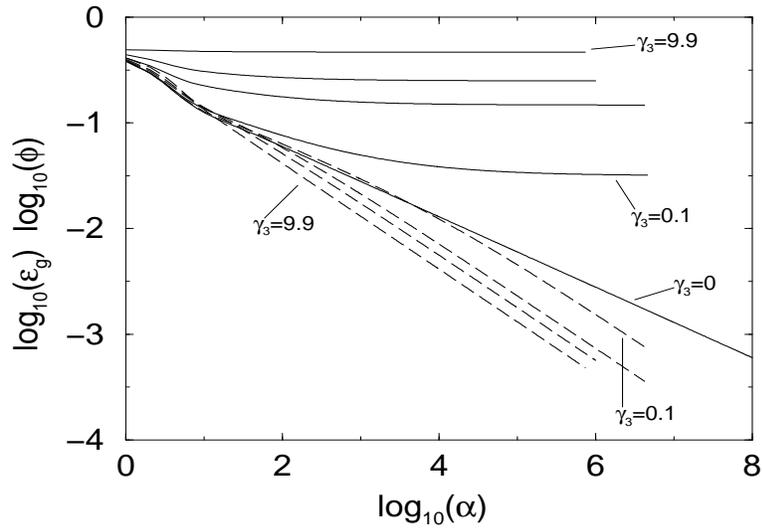}}
\caption{As in fig. \protect\ref{batp2batppm1} but for the perceptron
algorithm. Intermediate curves not marked on the figure are
for $\gamma_3=0.5,1$}
\label{batp2batppm2}
\end{figure}

\begin{figure}[h]
\epsfysize=8cm
\epsfxsize=10cm
\centerline{\epsfbox{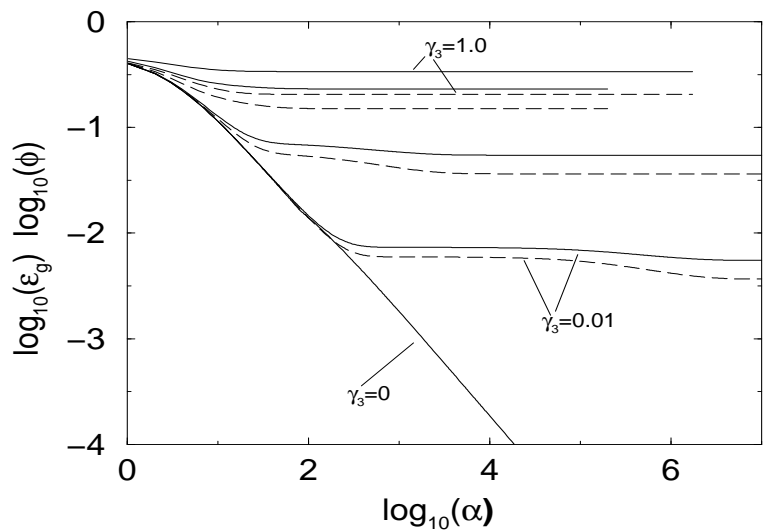}}
\caption{As in fig. \protect\ref{batp2batppm1} but for the Adatron
algorithm. Intermediate curves not marked on the figure are
for $\gamma_3=0.1,0.5$. }
\label{batp2batppm3}
\end{figure}

\begin{figure}[h]
\epsfysize=8cm
\epsfxsize=10cm
\centerline{\epsfbox{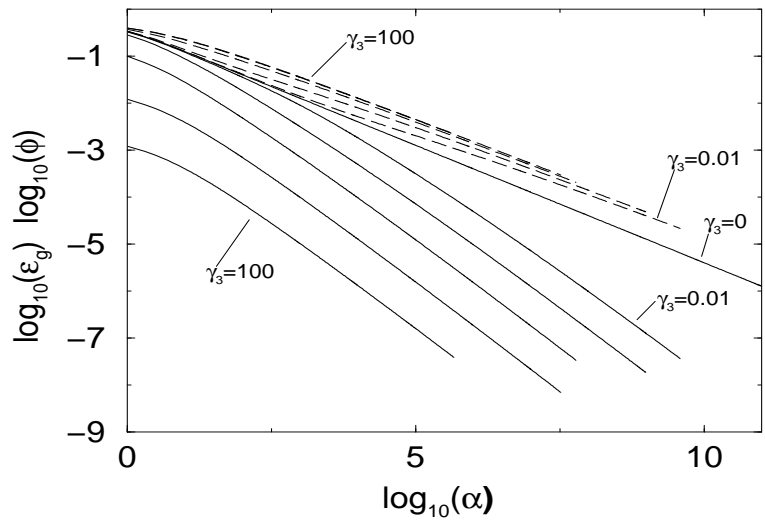}}
\caption{Learning scenario II: Log-log plot of  $\varepsilon_g$ (solid lines) 
and  $\phi$ (broken line), for $P=P_+$ and Hebbian learning as a
function of $\alpha$.
Intermediate curves not marked on the figure are 
for $\gamma_3=0.1,1.0,9.9$.}
\label{batp2batpp1}
\end{figure}

\begin{figure}[h]
\epsfysize=8cm
\epsfxsize=10cm
\centerline{\epsfbox{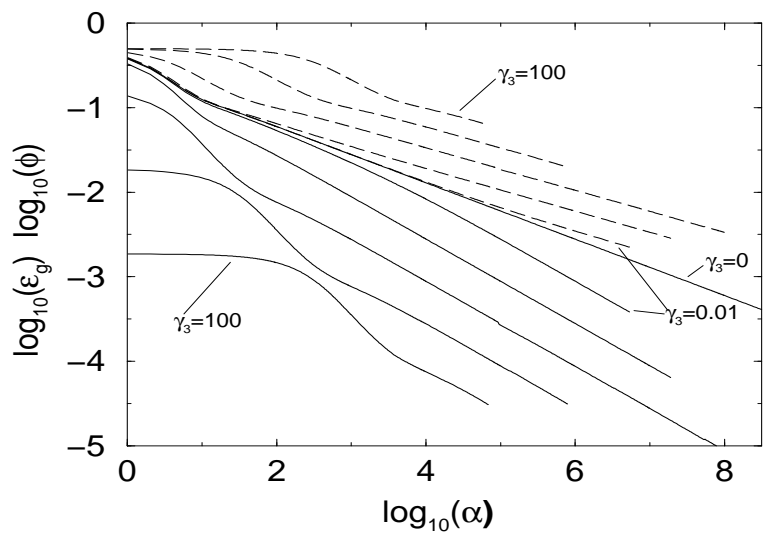}}
\caption{As in fig. \protect\ref{batp2batpp1} but for the perceptron
algorithm.}
\label{batp2batpp2}
\end{figure}

\begin{figure}[h]
\epsfysize=8cm
\epsfxsize=10cm
\centerline{\epsfbox{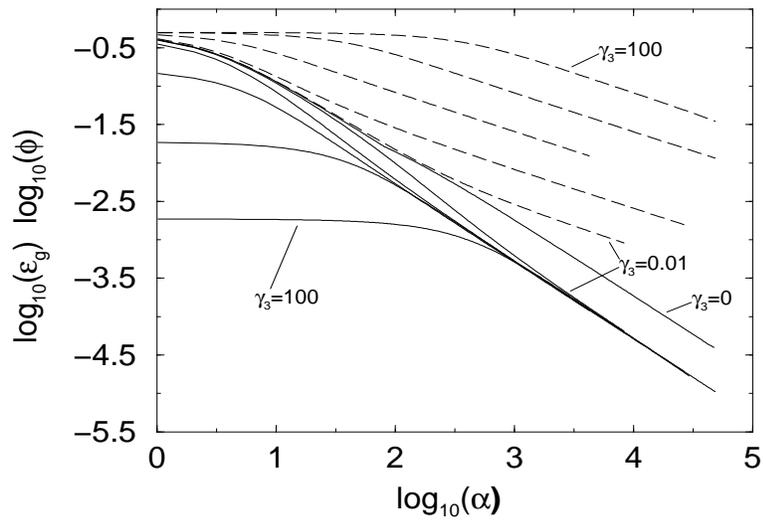}}
\caption{As in fig. \protect\ref{batp2batpp1} but for the Adatron
algorithm.}
\label{batp2batpp3}
\end{figure}

\begin{figure}[h]
\epsfysize=7cm
\epsfxsize=10cm
\centerline{\epsfbox{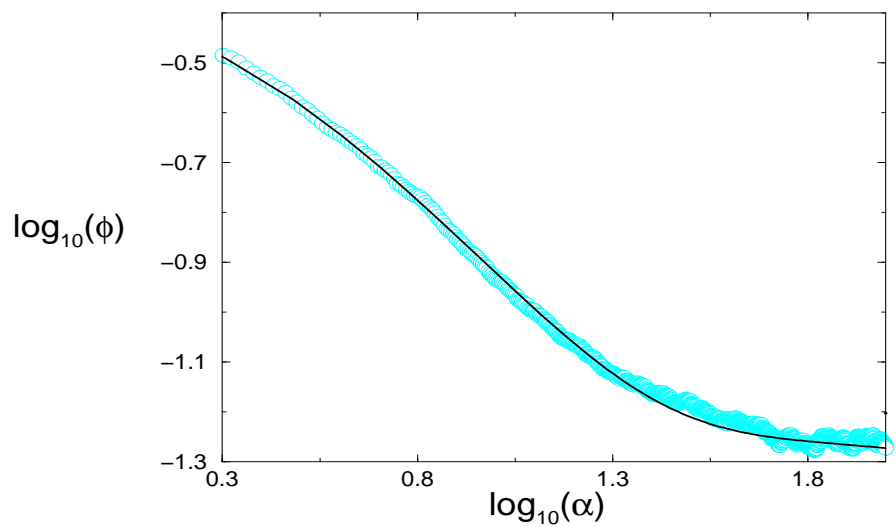}}
\caption{Second learning scenario with Adatron learning and $\gamma_3=1$. 
Simulations (grey circles) with $N=999$ versus theoretical
results (solid black line) for $\phi$ as  a function of $\alpha$.}
\label{simad2}
\end{figure}

\begin{figure}[h]
\epsfysize=7cm
\epsfxsize=10cm
\centerline{\epsfbox{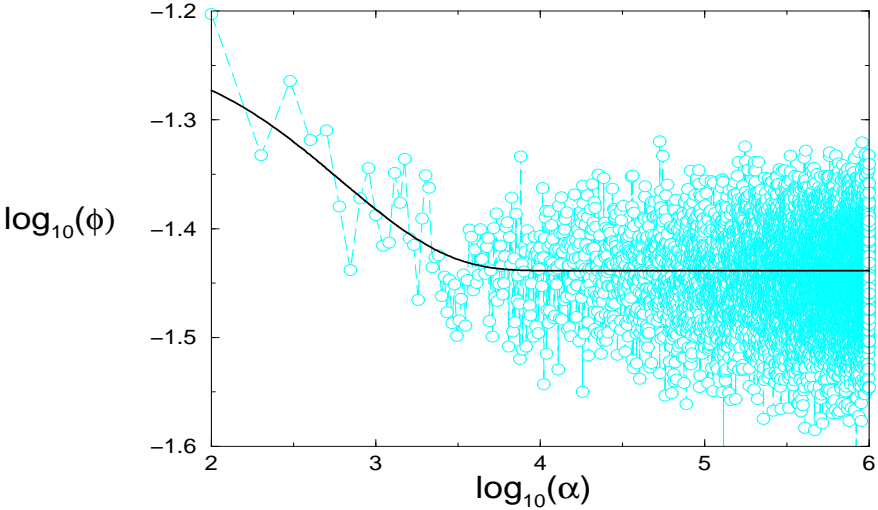}}
\caption{As in fig.\protect\ref{simad2}  with $N=100$.}
\label{simad2bis}
\end{figure}

\end{document}